\documentclass[aps,pra,twocolumn,longbibliography]{revtex4-2}

\usepackage{graphicx}
\usepackage{braket}
\usepackage{amsmath}
\usepackage{amssymb}
\usepackage{amsthm}

\usepackage[labelfont=bf]{caption}

\usepackage{float}
\usepackage{hyperref}

\begin{document}

\title{Multi-mode free-space delay interferometer with no refractive compensation elements for phase encoded QKD protocols}
\author{V. V. Tretiakov$^{1}$} \email{tretiakov.vv18@physics.msu.ru}
\author{K. S. Kravtsov$^{2,3}$}
\author{A. N. Klimov$^{1}$}
\author{S. P. Kulik$^{1}$}
\affiliation{
$^1$ Quantum Technology Centre of Moscow State University, Moscow, Russia\\
$^2$ Quantum Research Center, Technology Innovation Institute, Abu Dhabi, UAE\\
$^3$ Faculty of Physics, HSE University, Moscow, Russia
}

\date{\today}
\begin{abstract}
We demonstrate compensation-free approach to the realization of multi-mode delay interferometers, mainly for use in phase encoded quantum key distribution (QKD). High interference visibility of spatially multimode beams in unbalanced Michelson or Mach-Zehnder interferometers with a relatively wide range of delays is achieved by the appropriate choice of the transverse size of the beam. We provide a simple theoretical model that gives a direct connection between the visibility of interference, the delay, and the beam parameters.
The performed experimental study confirms our theoretical findings and demonstrates the measured visibility of up to 0.95 for the delay of 2 ns.
 Our approach's simplicity and robust performance make it a practical choice for the implementation of QKD systems, where a quantum signal is received over a multimode fiber. The important application of such configuration is an intermodal QKD system, where the free-space atmospheric communication channel is coupled into a span of the multimode fiber, delivering the spatially distorted beam to the remote receiver with minimal coupling loss.
\end{abstract}
\maketitle



\begin{center}
    \section{Introduction}
\end{center}

Quantum key distribution (QKD)~\cite{BEN84} or quantum cryptography is a method of sharing secret keys between two legitimate parties involved in a communication process.
Quantum cryptography is by far the most mature direction in the field of quantum technologies.  There are numerous commercial QKD systems available on the market.
The exceptional property that QKD relies upon is the available proofs of unconditional security for underlying protocols. Thus, its security is essentially guaranteed by the fundamental laws of quantum mechanics.
In this sense QKD is considered to be superior than conventional key exchange protocols, such as RSA and Diffie-Hellman~\cite{Diffie-Hellman} that only offer conditional security with no complete security proofs.

Efficient practical implementation of QKD is a significant challenge due to the requirement to perform measurements of light fields at a single photon level.
Until recently, most efforts were directed towards the implementation of QKD protocols over optical fiber infrastructure using different approaches and protocols, e.g. time-bin encoding \cite{Boaron-2018-time-bin}, COW \cite{Stucki_2009-cow}, entanglement-based protocols \cite{Wengerowsky_2019-entanglement}, MDIQKD \cite{PhysRevLett.117.190501-mdiqkd}, etc. That resulted in the development of whole metropolitan quantum networks \cite{Tokyo-network, Hefei-2010-network}. However, there are rather strict limitations on the achievable QKD distance over optical fibers due to the inevitable exponentially scaling signal attenuation. Therefore, originally weak QKD signals at some point become comparable to the intrinsic single photon detector noise, and the key exchange becomes impossible.

The alternative practical QKD implementation approach is based on free-space communication channels~\cite{Takenaka_2017-50-kg-sat, qkd-airplane, airborne-qkd, kravtsov2018relativistic}. First, it may serve as a viable last-mile option for servicing mobile users or areas without stationary optical fiber infrastructure. Second, it can provide global QKD connectivity by reaching to low orbit satellites and using them as trusted nodes~\cite{Liao_2017-satellite, satellite-entenglement}.
The majority of free-space QKD implementations uses polarization encoding protocols~\cite{Takenaka_2017-50-kg-sat, Liao_2017-satellite}, as they are simple to implement, and polarization states are virtually not affected by the propagation in the atmosphere~\cite{Hohn:69}.

For practical uses it may be convenient to separate the free-space channel terminals, typically located outside, and the actual QKD prepare and measure equipment, which by security considerations would rather be located inside, preferably in a ``trusted zone'' accessible exclusively by the authorized personnel.
Practical realization of this idea is straightforward at the transmitter side, where the quantum transmitter may be easily connected to the free-space coupling telescope via single-mode optical fiber (SMF).

The receiving side solution is a much more challenging one.
The use of an SMF for connecting the receiving telescope with the QKD receiver is feasible, but typically leads to large and time-varying coupling losses due to the accumulated atmospheric beam distortions.
By contrast, the use of multi-mode fibers (MMFs) allows one to avoid the coupling loss, but at the price of totally destroyed polarization states.
Thus, polarization encoding free-space QKD systems have essentially two options:
a) perform polarization state measurements at the receiving telescope, which is located somewhere outside; thus, secret keys appear not where they are used, but in a different location, so the problem of their secure delivery to, e.g. server room inside the building becomes an additional obstacle or even a security threat; b) deliver quantum states to the convenient location via the SMF, but suffer some 20~dB of coupling loss~\cite{padova, Avesani_2021}. As one can see, neither option is very appealing.


A real alternative is implementation of phase encoding QKD protocols~\cite{PhaseEncSMZI}, where quantum information is represented by the varying relative phase between temporally separated parts of an optical pulse.
A typical measurement scheme for receiving phase-encoded quantum states comprises a delay interferometer.
It is inherently challenging to design a spatially multi-mode delay interferometer with high interference visibility.
To the best of our knowledge, all previous efforts~\cite{Jin_2018, glass-tubes, Jin_2019, time-bin-4f, Castillo:23} relied upon the use of some refractive compensation elements that effectively worked as optical relays, reconstructing the delayed multi-mode beam to enable high quality interference.

In the present work we demonstrate that basic  Michelson or Mach-Zehnder delay interferometers without any diffraction compensation may still provide high interference visibility, despite the previously expressed doubts~\cite{Jin_2018}.
We show that high multi-mode visibility may be achieved by an appropriate choice of the transverse beam size, and the deterioration of visibility due to diffraction for large enough beams becomes negligible.
First, we provide a simple numerical model that can be used for simulation of interference in non-compensated multi-mode delay interferometers.
Second, we perform experimental measurements with multi-mode beams in a delay interferometer to confirm our theoretical findings.
The two approaches demonstrate reasonable agreement, and prove that compensation-free interferometers are totally feasible for MMF-connected receivers requiring the delay of the order of a few nanoseconds.

\section{Multi-mode free-space iterferometer modeling}
\begin{figure}[h]
    \centering
    \includegraphics[width=\linewidth]{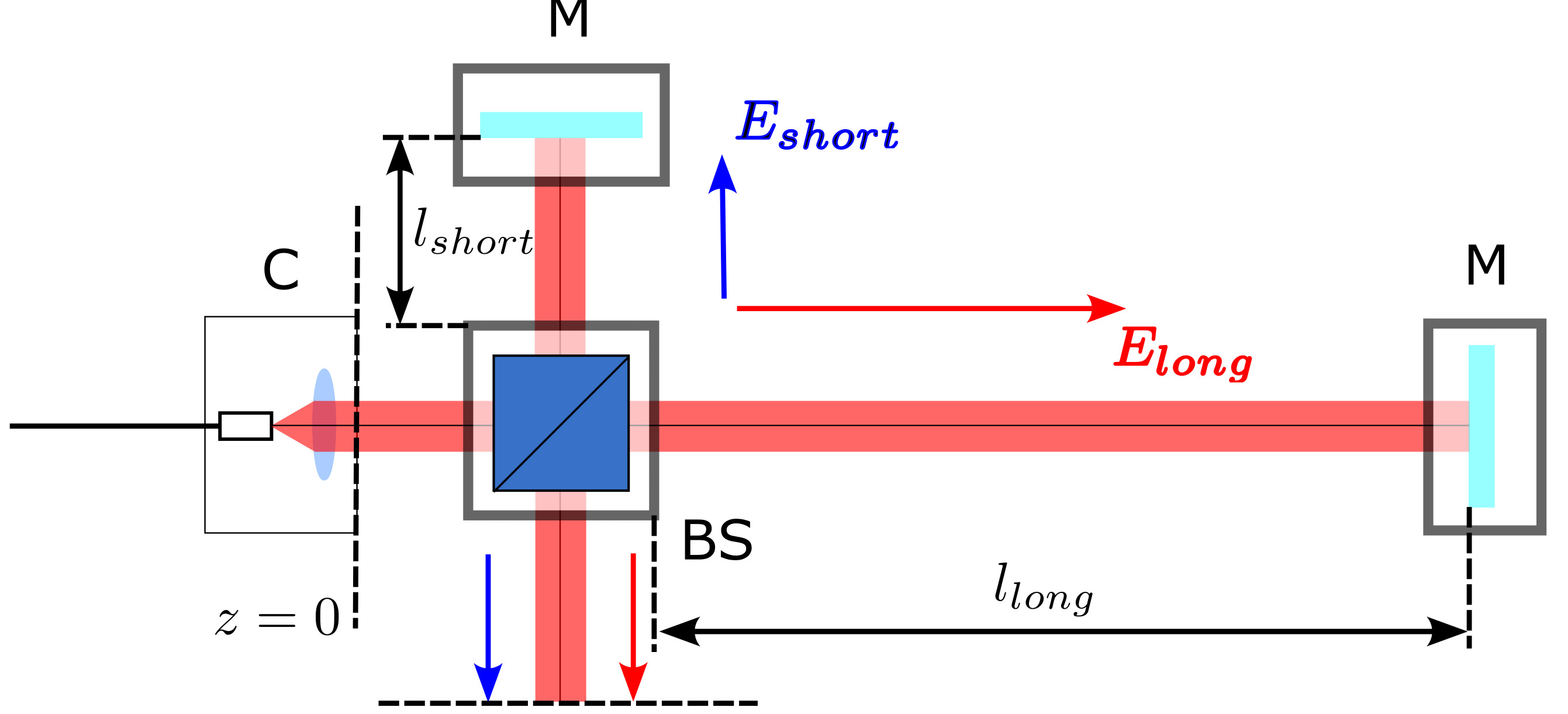}
    \caption{Michelson interferometer; C - collimator, M- mirror, BS - beamsplitter. The interference visibility at its output is calculated on the basis of evaluated field distribution in beams passing through the interferometer's arms.}
    \label{fig:delay_int}
\end{figure}

Consider a Michelson delay interferometer with the optical path difference of $\Delta z$ depicted in Figure~\ref{fig:delay_int}. Let plane $z = 0$ correspond to the collimator output, where the multimode collimated beam first appears. Our goal is to analyze the interference visibility at the output plane. As the interferometer arms are of different lengths, beam diffraction is also unequal; therefore, the visibility cannot be perfect. 

Following the paraxial approximation for the monochromatic wave, the electric field is given by:
\begin{equation}
    E(x, y, z, t) = A(x, y, z) e^{ikz} e^{i\omega t}, 
\end{equation}
where the complex field amplitude $A(x, y, z)$ can be decomposed in the orthogonal basis of Hermite-Gaussian modes:
\begin{equation} \label{eq:mode_decomp}
    A\left(x,y,z\right)=\sum_{\substack{n, \; m: \\ n+m < N}}{A_{nm}\left(x,y,z\right)} e^{i \phi_{nm}},
\end{equation}
where N is a combined order of the highest mode included in the field decomposition, $\phi_{nm}$ is a relative phase of the mode with the index $nm$.
For reasonable representation of typical multi-mode beams we use equal amplitudes for all included modes, while the phases are uniformly randomized in the
interval $(0, 2\pi)$.
Each Hermite-Gaussian mode is given by
\begin{equation}
    \begin{aligned} \label{eq:anm}
        A_{nm}\left(x,y,z\right)=\sqrt{\frac{\frac{2}{\pi}}{2^{n+m}n!m!}}\frac{1}{\sqrt{1+z^2}}H_n\left(\frac{\sqrt2x}{\sqrt{1+z^2}}\right)\times \\
        \times H_m\left(\frac{\sqrt2y}{\sqrt{1+z^2}}\right)\exp\left(-\frac{x^2+y^2}{1+z^2}\right)\exp\left(-i\frac{x^2+y^2}{z+\frac{1}{z}}\right)\times  \\
        \times \exp\Bigl(i\left(n+m+1\right)\arctan\left(z\right)\Bigr),
\end{aligned}
\end{equation}
where for simplicity $x$ and $y$ coordinates are measured in units of $w_0$ that defines the beam width, while $z$ coordinate is measured in the units of Rayleigh length $z_R = \frac{\pi w_0^2}{\lambda}$.

To assign a proper value to the highest mode order $N$ we assume that the number
of modes in the incident beam is limited by the MMF, connecting the receiving free-space coupler and the interferometer. The fiber is characterized by its
so-called V-parameter
\begin{equation}
    V = \frac{2\pi}{\lambda} \; \mathit{NA} \; a,
\end{equation}
where \(\lambda - \) the radiation's wavelength, \(\mathit{NA} - \) the optical fiber's numerical aperture, \(a -\) its core radius. \\

\begin{figure}[h]
    \centering
    \includegraphics[width=\linewidth]{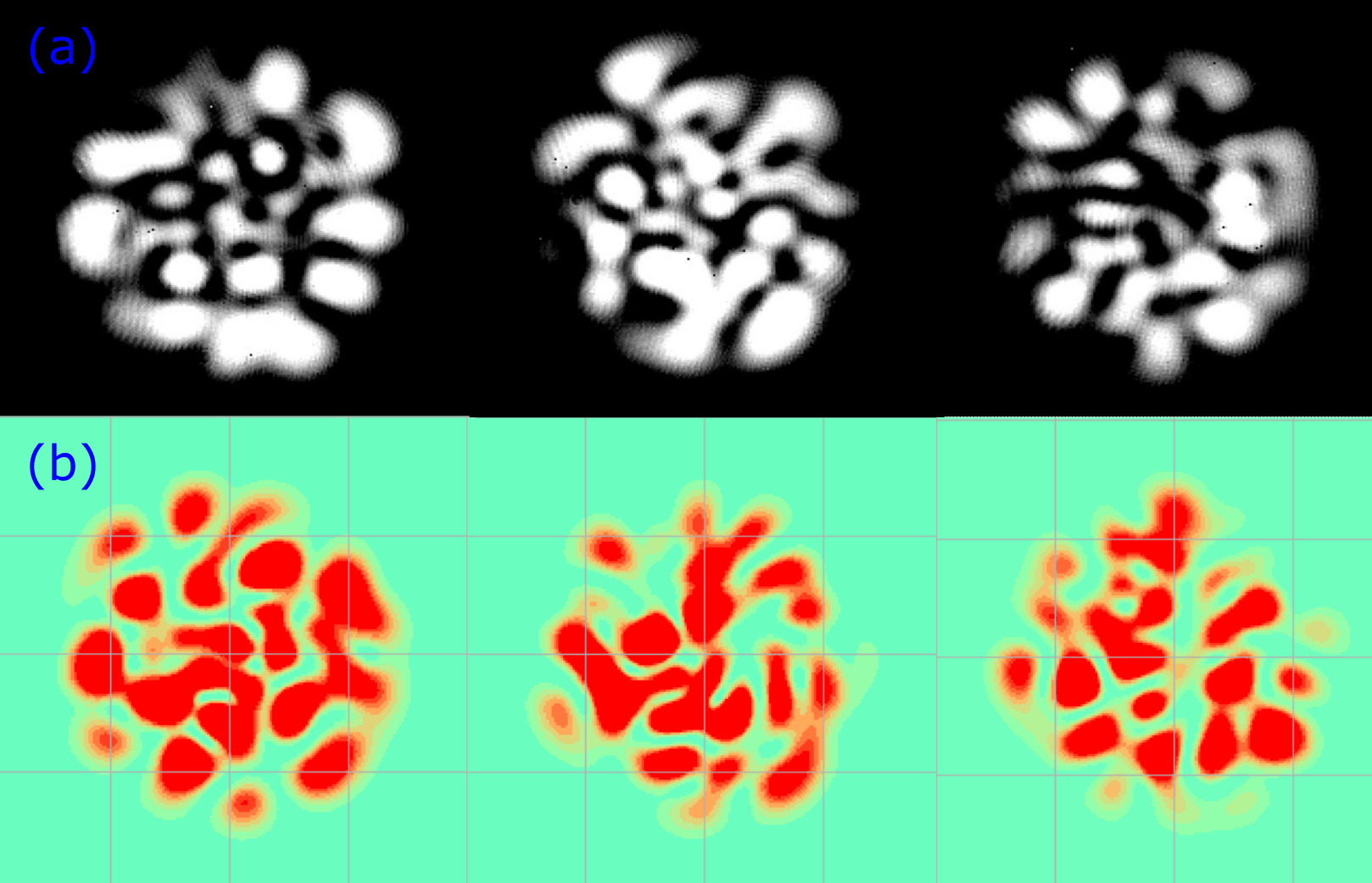}
    \caption{Typical intensity distribution in (a) experimentally measured collimated MMF outputs, and (b) in simulated beams with the highest mode order $N = 10$. The three columns differ by the random choice of relative phases.}
    \label{fig:MM_beams}
\end{figure}

It is known \cite{Saleh-Teich} that the total number of guided spatial modes (without considering polarization diversity) in the MMF with a parabolic refractive index profile, which we use in our experiment, is given by
\begin{equation}
    M \approx \frac{V^2}{8}.
\end{equation}
This also defines the parameter $N$, as $M= \frac{N(N+1)}{2}$. Parameters of our experiment are as follows: $\lambda = 1550 \; nm$, $\mathit{NA} = 0.2$, $a = 25 \; \mu m$, so $N \approx 10$.

The decomposition \eqref{eq:mode_decomp} provides an opportunity to find the total electric field distribution at arbitrary $z$.
To compare the chosen modal decomposition with the experiment, in Figure~\ref{fig:MM_beams} we show exemplary intensity cross-sections of the simulated multi-mode beams at $z=0$ and the experimentally measured ones. They demonstrate reasonable qualitative agreement.


In order to find the achievable multi-mode visibility we define the field cross-sections at the interferometer output that correspond to the two arms:
$\alpha (x, y) = A(x, y, z_1)$ and $\beta (x, y) = A(x, y, z_2)$, where $z_1$ and $z_2$ are beam propagation distances through the short and the long arms respectively. Thus, the total output field can be written as:
\begin{equation}
    E_{out}(x,y,z_1,z_2,t)=\alpha\left(x,y\right)e^{ikz_1}e^{i\omega t}+\beta\left(x,y\right)e^{ikz_2}e^{i\omega t}.
\end{equation}
As far as the intensity $I = E E^*$, the following expression for the intensity in the output plane can be obtained:
\begin{equation}
    \begin{aligned}
        I\left(x,y,z_1,z_2,\Delta z\right)=\left|\alpha\left(x,y\right)\right|^2+\left|\beta\left(x,y\right)\right|^2+ \\
        +2Re\left(\beta\left(x,y\right)\alpha^\ast\left(x,y\right)e^{ik\Delta z}\right).
    \end{aligned}
\end{equation}
Varying $\Delta z = z_2 - z_1$ within the limits of the wavelength and integrating the intensity over the whole plane to obtain the power $P\left(z^\ast,\Delta z\right)=\iint I\left(x,y,z^\ast,\Delta z\right)dxdy$, which can be detected, we obtain the expression for the interference visibility in a form:
\begin{equation} \label{eq:vis}
    \begin{aligned}
        V(z_1, z_2)=\frac{P^\mathrm{max}-P^\mathrm{min}}{P^\mathrm{max}+P^\mathrm{min}}= \\
        =\frac{2\left|{\displaystyle\iint}\beta\left(x,y\right)\alpha^\ast\left(x,y\right)dxdy\right|}{{\displaystyle\iint}\left(\left|\alpha\left(x,y\right)\right|^2+\left|\beta\left(x,y\right)\right|^2\right)dxdy}
    \end{aligned}
\end{equation}
Thus, to estimate the visibility at the output of the free-space delay interferometer we only need to calculate complex amplitude of the fields passed through the short and long arm of the interferometer, that we can do using the expression \eqref{eq:mode_decomp}.

\begin{figure}[h]
    \centering
    \includegraphics[width=\linewidth]{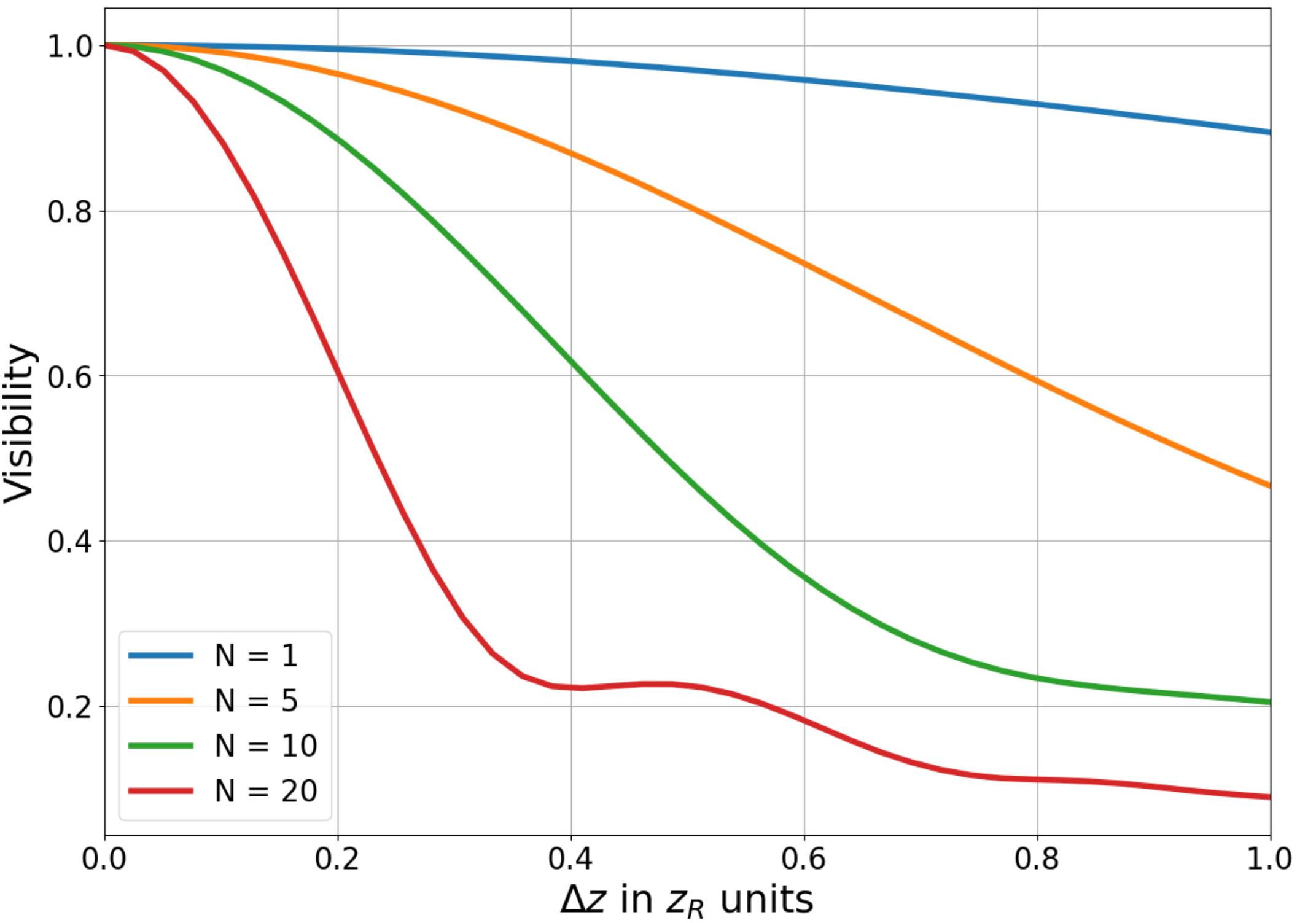}
    \caption{Simulated dependence of the multi-mode interference visibility as a function of the interferometer path length difference in $z_R$ units. The decrease of the visibility as a function of delay is explained by the beam diffraction. The more modes the faster the diffraction pattern evolves, resulting in poorer visibility.}
    \label{fig:MM_vis_z}
\end{figure}

As far as $z_1$ and $z_2$ define the delay in the interferometer (assuming that $z_1 = 0$ for the following numerical experiment), we obtain the dependence of the multi-mode interference visibility as a function of delay. 
Due to the inherently random choice of phases for different spatial modes, simulations yield slightly different results. In Figure~\ref{fig:MM_vis_z} we show the averaged results of 20 different numerical experiments.

\begin{figure}[h]
    \centering
    \includegraphics[width=\linewidth]{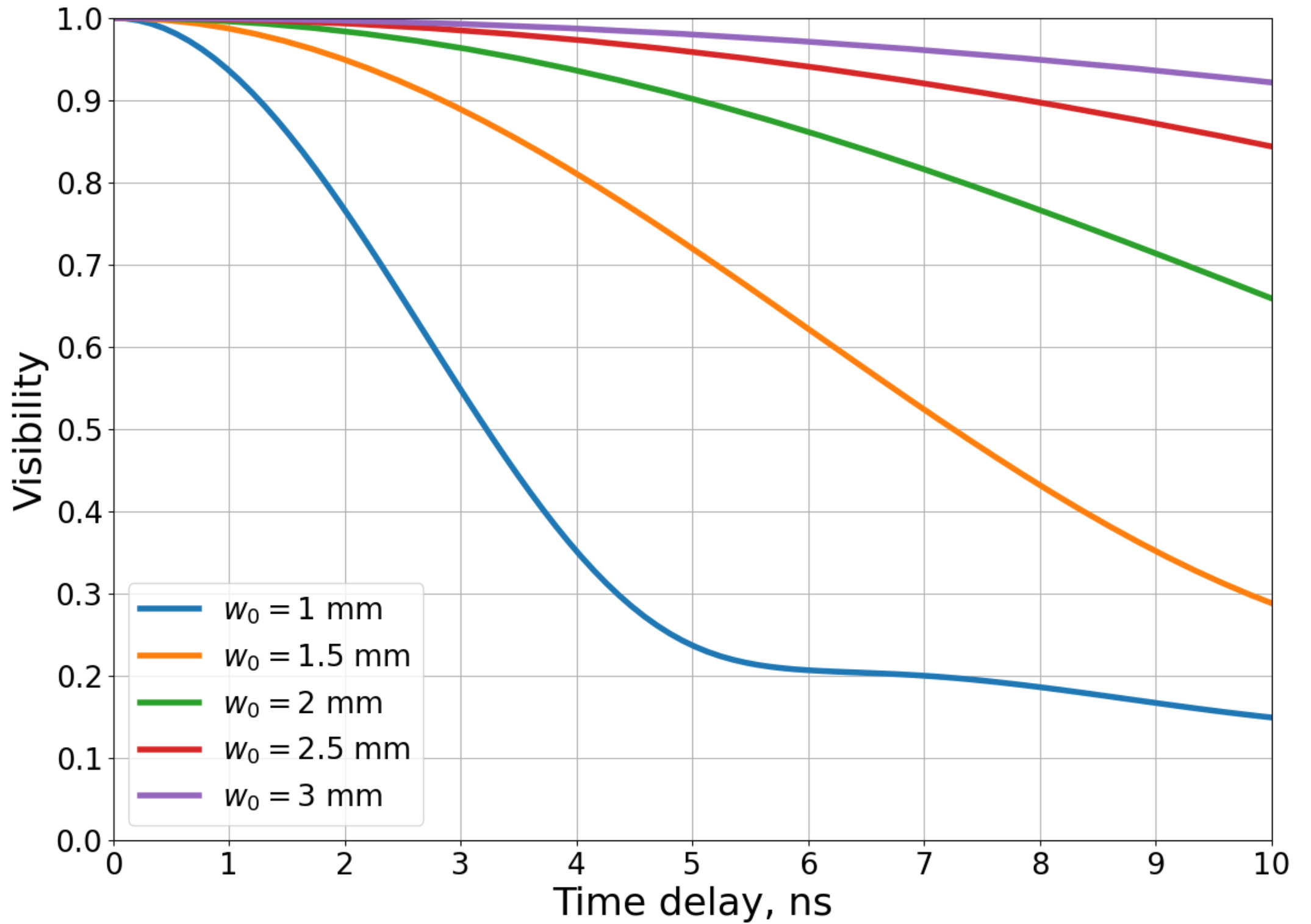}
    \caption{Theoretical dependence of the multi-mode interference visibility on the delay in the Michelson interferometer. Curves represent this dependence for different spot size parameters of the beam with $N = 10$. Since beam divergence is inversely proportional to the spot size parameter, theoretically achievable visibility for beam with smaller $w_0$ is reduced comparing to the beam with greater $w_0$.}
    \label{fig:MM_vis_t}
\end{figure}

So far we demonstrated that a reasonable choice of the Rayleigh range for the particular interferometer delay yields almost perfect interference visibility, which is of paramount importance for implementation of phase-encoding QKD protocols.
To assist in converting the rather abstract units to actual delays and beam radii, in Figure~\ref{fig:MM_vis_t} we show the dependence of visibility as a function of the delay for $N=10$ and different $w_0$.

We underline that the obtained results need to be treated as ideal, theoretically achievable. They inevitably slightly overestimate visibility values one can obtain in the experiment due to imperfections either in a collimation system causing the wave front aberrations or in other optical elements presented in the interferometer. Nevertheless, these results are very important since they show that theoretically it is possible to achieve almost perfect interference for the wide range of delays even in the case of a multi-mode signal at the input of the interferometer with no diffraction compensation.

\begin{figure}
    \centering
    \includegraphics[width=\linewidth]{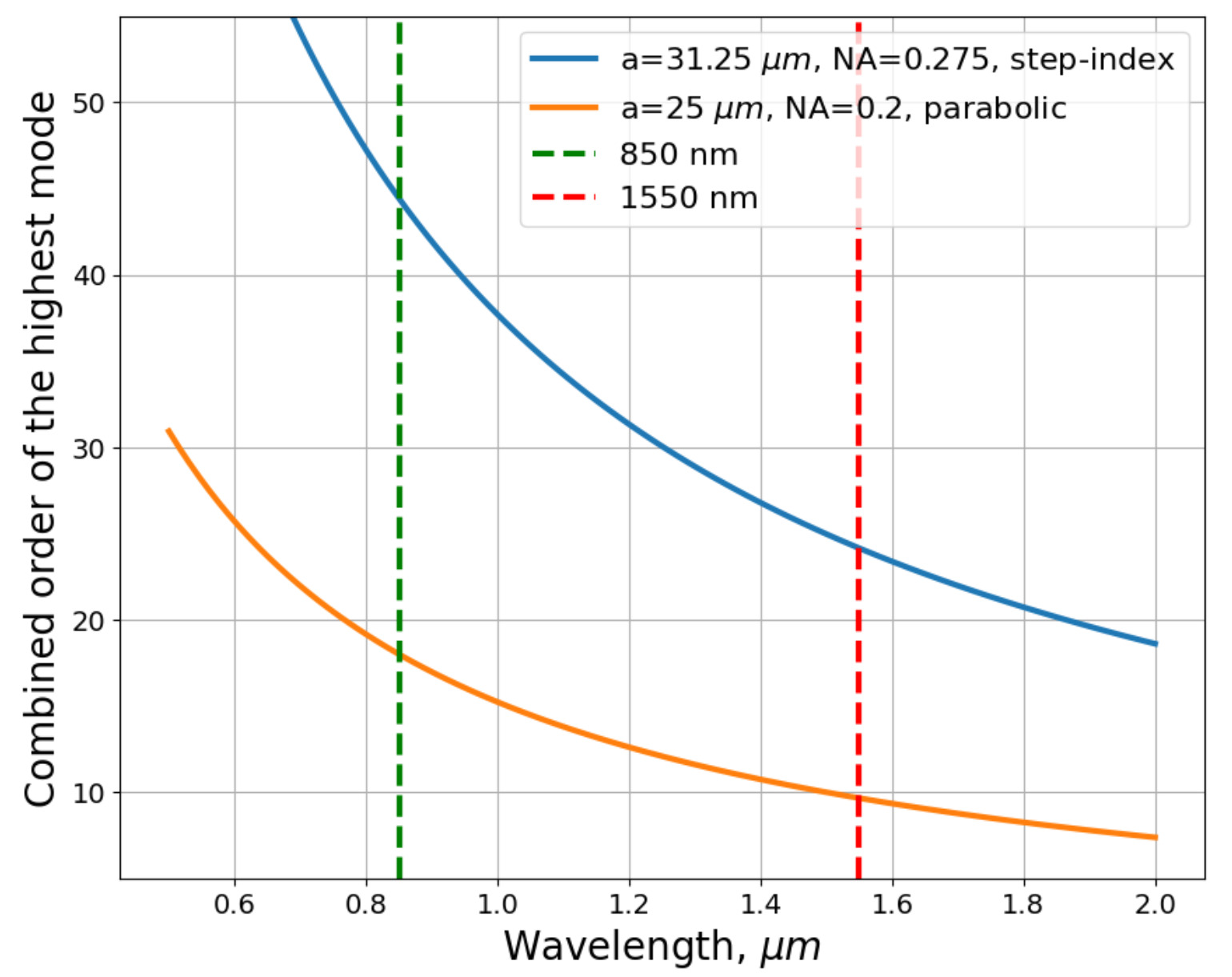}
    \caption{The the highest mode order $N$ as a function of the wavelength for two different MMFs with specified parameters. The wavelengths of 850~nm and 1550~nm are common choices for free-space QKD systems.}
    \label{fig:N_vs_lambda}
\end{figure}

\begin{figure}[h]
    \centering
    \includegraphics[width=\linewidth]{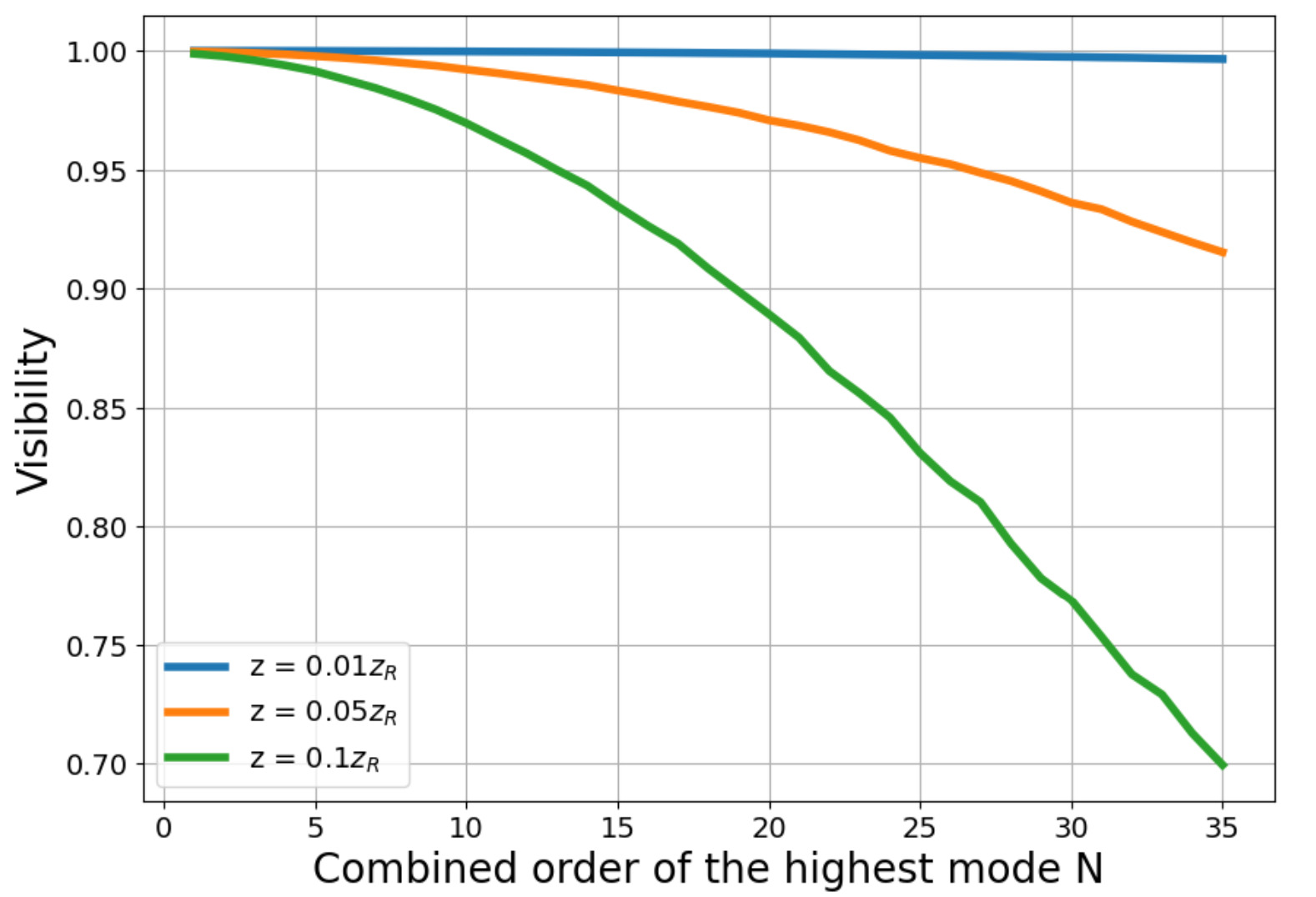}
    \caption{Theoretically achievable visibility as a function of the highest mode order $N$ for a given delay in $z_R$ units.}
    \label{fig:vis_vs_N}
\end{figure}

The number of modes in the incident multi-mode beam is limited by the MMF and strongly depends on its parameters, namely the numerical aperture and core radius, as well as on the radiation wavelength.
Figure~\ref{fig:N_vs_lambda} explicitly shows the dependence of the highest mode order $N$ on the wavelength for two common types of the MMF. 
Figure~\ref{fig:vis_vs_N} shows the achievable visibility as a function of the highest mode order $N$ for three fixed delays.
This information may be readily used for the appropriate choice of the MMF type for a particular operation wavelength. Proper choice of $N$ is critically important in compensation-free multi-mode interferometers as it essentially defines the achievable interference visibility. 
In the upcoming section, we present the findings from our experimental study on the visibility of multi-mode interference to further support our simulation results.


\begin{center}
    \section{Interference visibility experiment}
\end{center}

\begin{figure}[h]
    \centering
    \includegraphics[width=\linewidth]{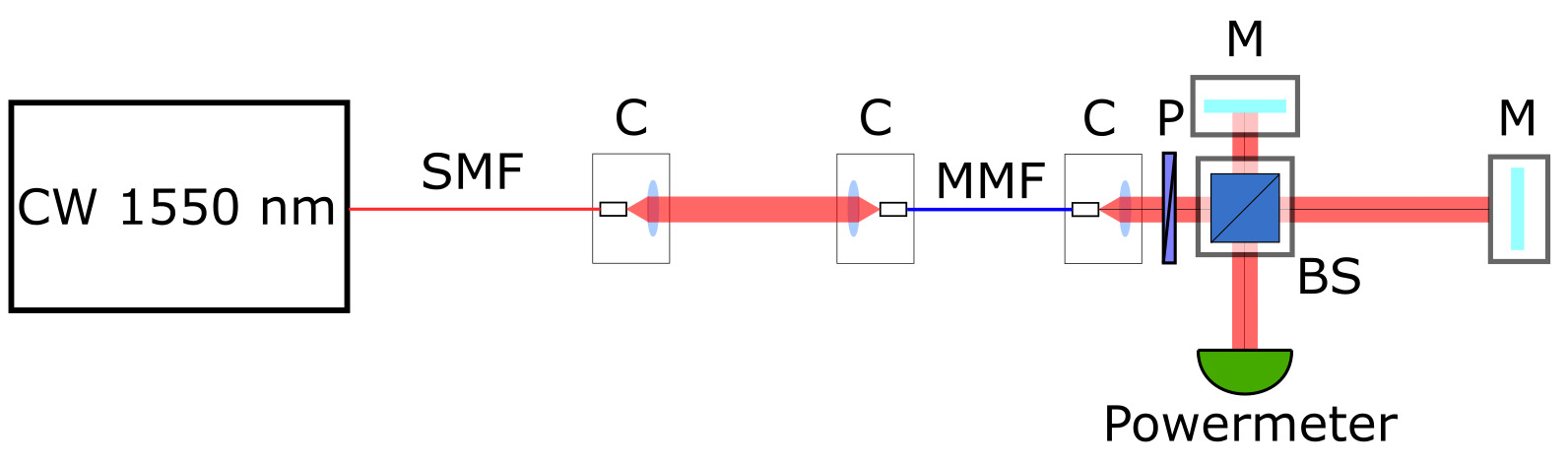}
    \caption{Schematic setup: C - collimator, P - polarizer, MMF/SMF - multi/single-mode fiber, BS - beamsplitter, M - mirror.}
    \label{fig:setup}
\end{figure}

In the performed experiment we used the setup shown in Figure~\ref{fig:setup}. We use a benchtop laser source Thorlabs TLX1, which provides continuous wave laser radiation at the wavelength of 1550 nm. Since the output of this source is single-mode whereas we investigate the multi-mode interference, it is necessary to couple the radiation into a multi-mode optical fiber such that all modes of this fiber are excited. For this purpose we use a 2-collimator system with a beam entering the MMF at some non-zero angle with the respect to the axis passing through the optical fiber core. Such the coupling regime results in a higher fiber modes excitation, typical radiation cross-section of which are depicted in Figure~\ref{fig:MM_beams}. The number of modes in an optical fiber defines the achievable interference visibility according to Figure~\ref{fig:MM_vis_z}. Thus, to see the worst-case scenario, we need to excite as many modes in a multi-mode fiber as possible.

After passing the 2-collimator system and 2-meter long MMF the radiation enters Michelson interferometer; we measure the power at the interferometer's output with a powermeter. To find the maximal and the minimal output power that defines the visibility we apply small controllable shifts of the order of $\lambda$ to one of the mirrors.

\begin{figure}[H]
    \centering
    \includegraphics[width=\linewidth]{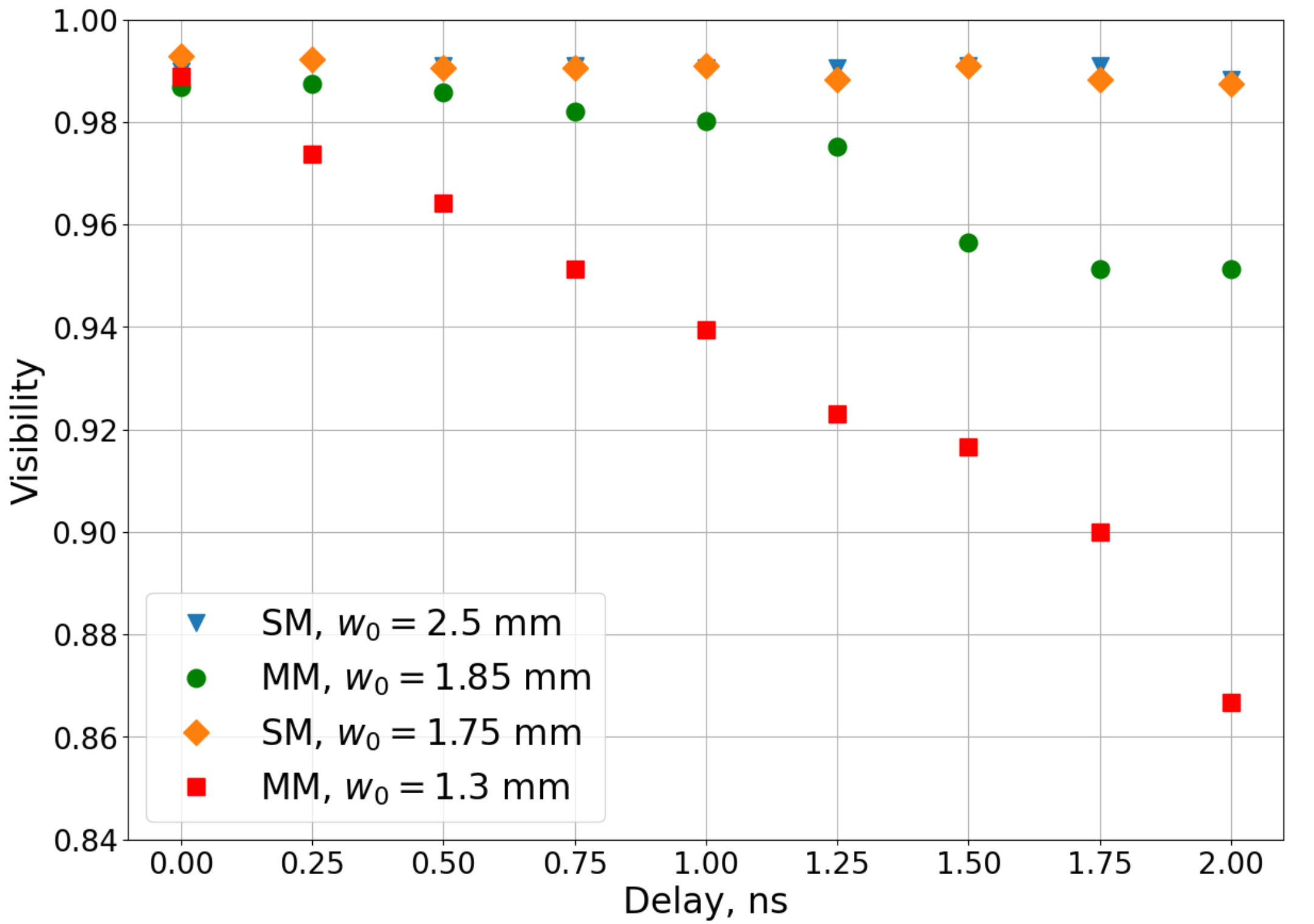}
    \caption{The interference visibility experiment results. SM/MM - single-mode and multi-mode beams, respectively, $w_0$ - spot size parameter.}
    \label{fig:Exp_vis}
\end{figure}

In order to obtain the dependence of the multi-mode interference visibility on the delay, we shift the mirror in the long arm with steps of 3.75 cm that corresponds to the delay variation of 0.25 ns. For the sake of repeatability we initially adjust the balanced Michelson interferometer. Then after each shift of the mirror in the long arm we only adjust this mirror leaving the other optical components untouched. As one of the purpose of this work is to show the dependence of the multi-mode interference visibility on the beam radius, we conduct the experiment for two different beam radii of 1.3~mm and 1.85~mm. The results are shown in Figure~\ref{fig:Exp_vis}. As expected, the visibility inevitably decreases with both the increase of the delay and the decrease of the spot size parameter.
For comparison we also plot the results of visibility measurements for single-mode Gaussian beams. They show uniform visibility with negligible deterioration with the delay.

\begin{figure}[h]
    \centering
    \includegraphics[width=\linewidth]{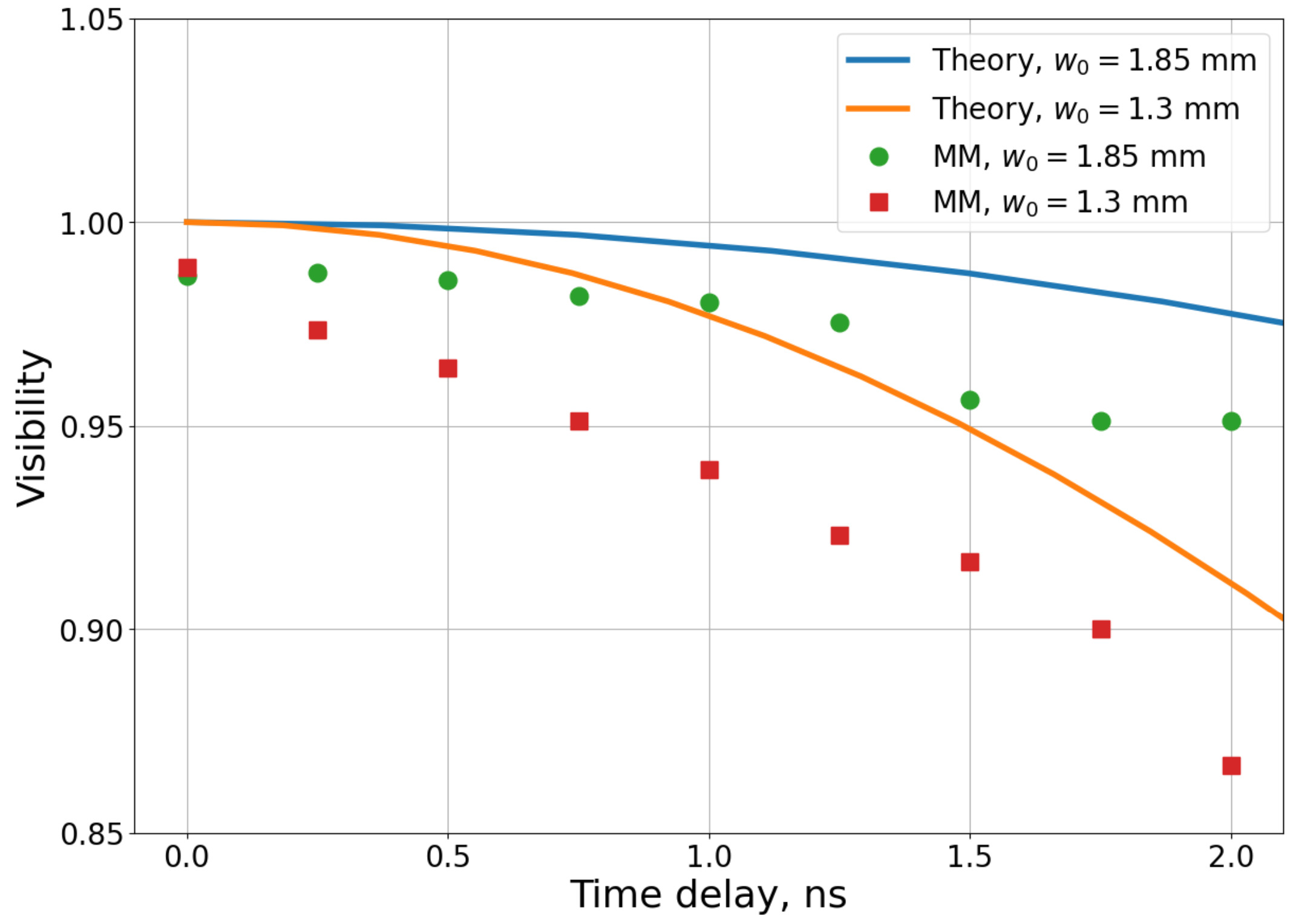}
    \caption{Comparison of interference visibility as a function of the delay between the experimental data (scatter plot) and the simulation results (solid lines).}
    \label{fig:theory_vs_exp}
\end{figure}

Figure~\ref{fig:theory_vs_exp} shows the comparison between the experimental data and the simulation results for our experimental parameters. One can see virtually the same behaviour with just a small amount of penalty for the experimentally observed visibility values, which come from the general experimental imperfections. Besides this penalty, both curves show fairly good agreement, thus confirming our theoretical findings.

\begin{figure}[h]
    \centering
    \includegraphics[width=\linewidth]{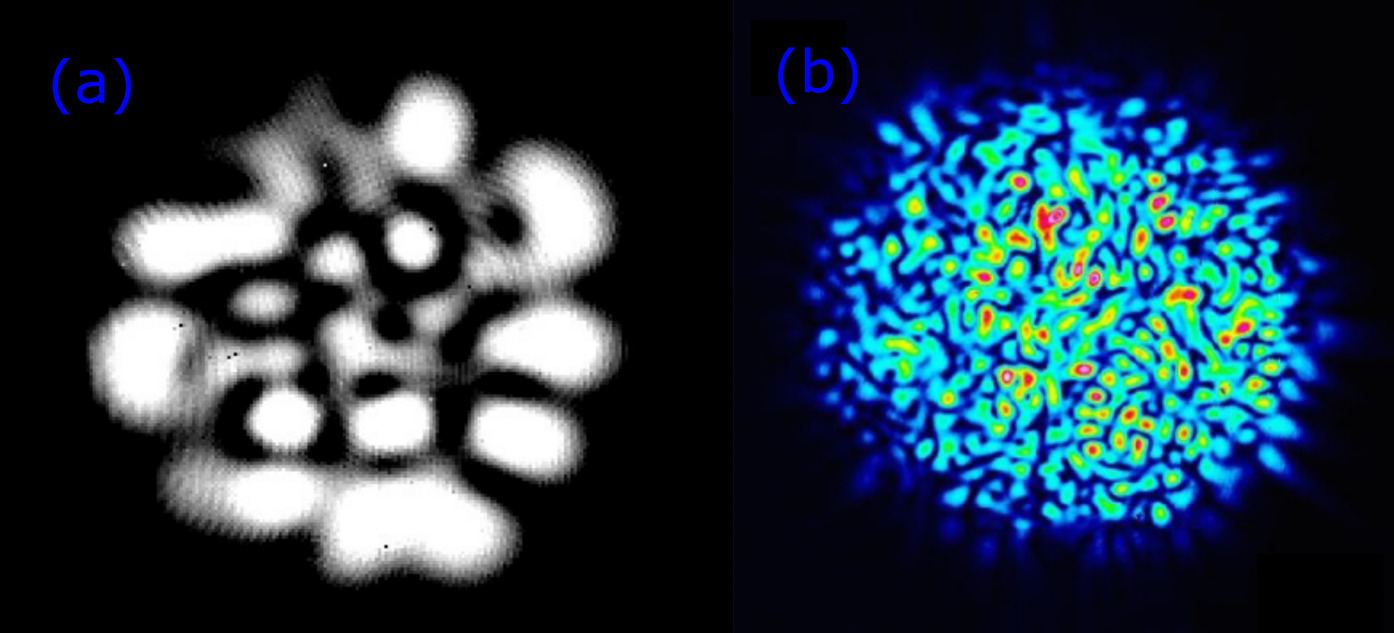}
    \caption{Typical cross-section intensity distribution in a beam passed through a) parabolic refractive index MMF with $a=25 \; \mu m, \; \mathit{NA} = 0.2$ and b) step-index MMF with $a=52.5 \; \mu m ;\ \mathit{NA} = 0.22$ \cite{Jin_2018}.}
    \label{fig:MM-beams_3}
\end{figure}

With regard to the fiber type selection, one can conclude that in terms of multi-mode interference visibility the appropriate MMF type is one with lower number of maintaining modes, e. g. lower core radius and numerical aperture. Also it is more preferable to choose a MMF with near-parabolic refractive index profile rather that step-index one that supports twice as large number of modes. Experimental results obtained by Jin \textit{et al} in \cite{Jin_2018} show extremely low values of multi-mode interference visibility without refractive compensation elements in delay interferometers. This fact is explained by the choice of the step-index MMF with $a = 52.5 \; \mu m, \; \mathit{NA} = 0.22$, what leads to the combined order of the highest mode $N = 60$. Figure~\ref{fig:MM-beams_3} shows typical multi-mode beams a) in our experiment and b) in \cite{Jin_2018}. As one can see, the choice of the MMF type and the operation wavelength significantly impacts the experimentally achievable visibility.

There is an apparent trade-off between the free-space coupling efficiency from the atmospheric communication line to the MMF, which is good for larger core fibers; and the interference visibility, which is better for smaller cores. An additional tuning parameter is the beam size $w_0$. In theory, it can be set to the large enough value to accommodate for any number of modes. However, increasing $w_0$ requires larger and more precise collimation optics. Even more importantly, larger beams lead to the increase of angular sensitivity of interferometer's mirrors, which makes it more demanding in tuning and ensuring long-term stability.
Therefore, a reasonable strategy could be to operate within a 'sweet spot' in this trade-off partially sacrificing the coupling loss to maintain the appropriate visibility.

\section{Conclusion}

We theoretically investigated the dependence of interference visibility in  a multi-mode delay interferometer on the number of modes, the beam size, and the delay. Our model revealed that given the beam radius and the delay are chosen appropriately, it is possible to observe almost perfect interference visibility in the delay interferometer without refractive compensation elements. The obtained experimental data proves that statement; therefore, the studied approach to realization of delay interferometers 
is suitable for the implementation of phase-encoding QKD protocols. The simplicity of this approach makes it a promising solution for the commercial implementation of QKD systems. It is straightforward and does not require technological sophistication. Our study also revealed that using the MMF with a lower numerical aperture and core radius between the receiving telescope and measurement device results in a smaller number of excited modes and, thus, higher interference visibility.


\bibliography{arxiv_sub}

\end{document}